\documentclass[12pt]{article}\usepackage[hyperfootnotes=false]{hyperref}
   \usepackage{epsfig}
      \usepackage{amsmath}
      \usepackage{amssymb}
      \usepackage{caption}

\usepackage{amsmath}

\usepackage{hyperref} 
\usepackage{setspace}
  \usepackage{graphicx}
  \usepackage{color}
 \newcommand{\be}{\begin{equation}}
\newcommand{\bea}{\begin{eqnarray}}
\newcommand{\eea}{\end{eqnarray}}
\newcommand{\beq}{\begin{equation}}
 \newcommand{\ee}{\end{equation}}
\def\tr{$\text{tr}$}

\def\nref#1{(\ref{#1})}

\def\la{\label}

\begin{document}
  \renewcommand{\theequation}{\thesection.\arabic{equation}}

\begin{titlepage}
  \rightline{NSF-KITP-15-162}
  \bigskip

  \bigskip\bigskip\bigskip\bigskip

  \bigskip
\centerline{\Large \bf {Relative entropy equals bulk relative entropy }}

    \bigskip

  \begin{center}

 \bf { Daniel L. Jafferis$^1$, Aitor Lewkowycz$^2$, Juan Maldacena$^3$, S. Josephine Suh$^{4,5}$ }
  \bigskip \rm
\bigskip

$^1${\it  Center for Fundamental Laws of Nature, Harvard University, Cambridge, MA, USA}
\smallskip

$^2${\it  Jadwin Hall,  Princeton University, Princeton, NJ, USA}
\smallskip 

 $^3$ { \it Institute for Advanced Study, 
 Princeton, NJ, USA\\ }
 
 \smallskip 
 
  $^4${\it  Kavli Institute for Theoretical Physics, Santa Barbara, CA, USA}
\smallskip 
 
 $^5${\it  Department of Physics and Astronomy, University of British Columbia, Vancouver, B.C., Canada }
\smallskip

\vspace{1cm}
  \end{center}

  \bigskip\bigskip

 \bigskip\bigskip
  \begin{abstract}

We consider the gravity dual of the modular Hamiltonian associated to a general subregion of a boundary theory. 
We use it to argue that the relative entropy of nearby states is given by the relative entropy in the bulk, 
to leading order in the bulk gravitational coupling. We also argue
that the boundary modular 
flow is dual to the bulk modular 
flow in the entanglement
wedge, with implications for entanglement wedge reconstruction.

 \medskip
  \noindent
  \end{abstract}

  \end{titlepage}

  \tableofcontents

\section{Introduction and summary of results}

Recently there has been a great deal of effort in elucidating patterns of entanglement for theories that 
have gravity duals.   The simplest quantity that can characterize such patterns is the von Neumann entropy of subregions, sometimes
called the ``entanglement entropy''. This quantity is divergent in local quantum field theories, but the divergences are  well understood and one 
can extract finite quantities. Moreover, one can   construct    strictly finite quantities  that 
are well-defined and have no ambiguities. 
A particularly interesting quantity is the so called ``relative entropy''  \cite{Ohya.Petz,Araki:1976zv}. 
This is a measure of distinguishability between two states, a reference ``vacuum state'' $\sigma$ and
an arbitrary state $\rho$
\be
 S(\rho|\sigma) = Tr[ \rho \log \rho - \rho \log \sigma ] 
 \ee
 If we define a modular Hamiltonian $ K = - \log \sigma$, then this can be viewed as the free energy difference between the state $\rho$ and the
 ``vacuum'' $\sigma$ at temperature $\beta =1$, $S(\rho|\sigma) = \Delta K - \Delta S $. 
 
 Relative entropy has nice positivity and monotonicity properties. It has also played an important role in formulating a precise version of the Bekenstein bound  
\cite{Casini:2008cr} and arguments for the second law of
 black hole thermodynamics  \cite{Wall:2010cj, Wall:2011hj}. 
 
 In some cases the modular hamiltonian has a simple local expression. The simplest case is the one associatated to Rindler space, where the modular Hamiltonian 
 is simply given by the boost generator.  
 
\begin{figure}[h!]
\begin{center}
\vspace{5mm}
\includegraphics[scale=1.25]{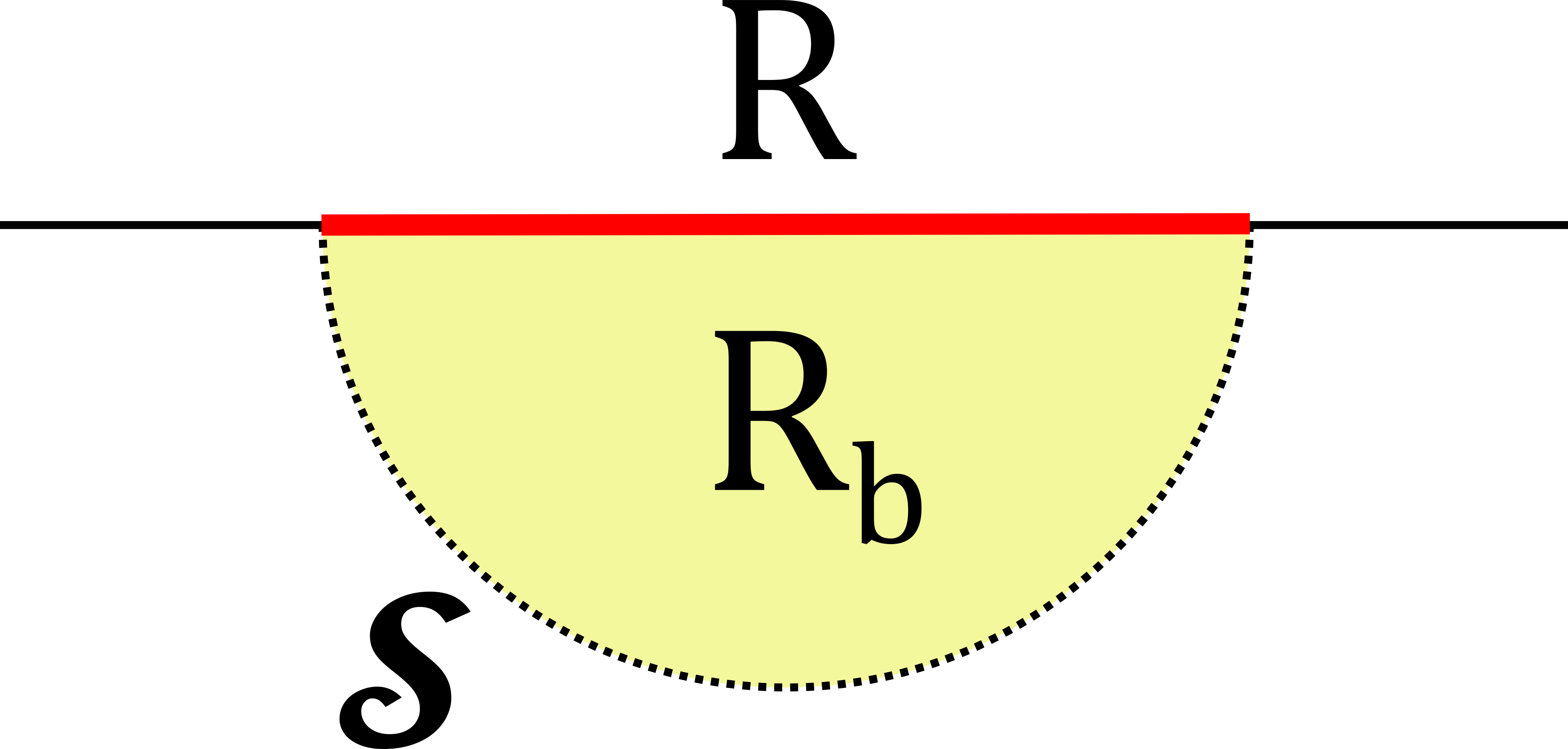}
\vspace{5mm}
\caption{
The red segment indicates a spatial region, $R$,  of the boundary
theory. The leading contribution to the entanglement entropy is computed by the area of an extremal surface
$\cal S$ that ends at the boundary of
region $R$. This surface divides the bulk into two, region $R_b$ and its complement. Region
$R_b$ lives in the bulk and has one more dimension than region $R$. The leading correction
to the boundary entanglement entropy
is given by the bulk entanglement entropy between region $R_b$ and the rest of the bulk.
 }
\label{RTBasic}
\end{center}
\end{figure}

 In this article we consider quantum field theories that have a gravity dual.  We consider an arbitrary subregion on the boundary theory $R$, and a reference state $\sigma$, 
 described by a smooth gravity solution. $\sigma$ can be the vacuum state, but is also allowed to be any state described by the bulk gravity theory. 
 We then claim that the modular Hamiltonian corresponding to this state has a simple bulk expression. 
 It is given by 
 \be \la{ModHam}
 K_{\rm bdy} = { {\rm Area}_{\rm ext} \over 4 G_N} + K_{\rm bulk } + \cdots + o(G_N) 
 \ee
 The first term is the area of the Ryu Takayangi surface ${\cal S}$ (see figure \ref{RTBasic}), viewed as an 
 operator in the semiclassically quantized bulk theory.
 This term was previously discussed in \cite{Jafferis:2014lza}.
   The $o(G_N^0)$ term $K_{\rm{bulk}}$ is the modular Hamiltonian 
 of the bulk region enclosed by the Ryu-Takayanagi surface, $R_b$, when we view the bulk as an ordinary quantum field theory, with suitable care exercised to treat the 
 quadratic action for the gravitons. Finally,  the dots represent  local operators on ${\cal S}$, which we will later specify.
  We see that the boundary modular Hamiltonian has a simple expression in the bulk. In particular, to leading order in the $1/G_N$ expansion it is just the area 
  term, which is a 
  very  simple local expression in the bulk. Furthermore, this simple expression is precisely what appears in the entropy. 
   This modular Hamiltonian makes sense when we compute its action on bulk field theory states $\rho$ which are related to  
     $\sigma$ by bulk perturbation theory. Roughly speaking, 
 we consider a $\rho$ which is obtained from $\sigma$ by adding or subtracting particles without generating a large backreaction. 
  
  Due to the form of the modular Hamiltonian \nref{ModHam}, 
  we obtain a simple result for the relative entropy
  \be \la{RelRel}
  S_{\rm bdy} (\rho|\sigma)  = S_{\rm bulk}(\rho | \sigma) 
  \ee
  where the left hand side is the expression for the relative entropy on the boundary. In the right hand side we have the relative entropy of the bulk 
  quantum field theory, with $\rho$ and $\sigma$ in the right hand side, being the bulk states associated to the boundary states $\rho, \sigma$ appearing in the
  left hand side. Note that the area term cancels. 
  
Another consequence of \nref{ModHam} is that the action of $K_{\rm bdy}$ coincides with the action of $K_{\rm bulk}$ in the interior of the entanglement wedge\footnote{The entanglement wedge is the 
domain of dependence of the region $R_b$.}, 
\be \label{Keq}
[ K_{\text{bdy}},\phi]=[K_{\text{bulk}},\phi]
\ee
for $\phi$ a local operator in $R_{b}$. 
This follows from causality in bulk perturbation theory: terms in $K_{\rm bdy}$ localized on ${\cal S}$ do not contribute to its action in the interior of the entanglement wedge, ${\cal S}$ being space-like to the interior. Note $K_{\rm bulk}$ is the bulk modular Hamiltonian associated to a very specific subregion, that bounded by the extremal surface ${\cal S}$. Implications of \eqref{Keq} for entanglement wedge reconstruction are described in section \ref{sec: EW}.

The bulk dual of relative entropy for subregions with a Killing symmetry   was considered before   in \cite{Blanco:2013joa,Lashkari:2013koa,Faulkner:2013ica,Lin:2014hva,Lashkari:2014kda,Lashkari:2015hha}. In particular, in  
\cite{Lashkari:2015hha}, the authors related it to the classical canonical energy. 
In fact, we argue below that the bulk modular hamiltonian is equal to the canonical energy in this case. 
This result extends that discussion to the quantum case. Note \nref{ModHam} and \nref{RelRel} are valid for arbitrary regions, with or without a Killing symmetry. In addition, we are not restricting
$\sigma$ to be the vacuum state. 
 Recently a different extension of  \cite{Lashkari:2015hha} has been explored in \cite{LLOvR},
   which extends it to situations where one has a very large deformation relative to the 
 vacuum state. That discussion does not obviously overlap with ours. 

This paper is organized as follows. 
In section two, we recall definitions and properties of entanglement entropy, the modular Hamiltonian, and relative entropy. 
In section three, we present an argument for the gravity dual of the modular hamiltonian and the bulk expression for relative entropy. 
In section four, we discuss the case with a $U(1)$ symmetry, relating to previous work. 
 In section five, we discuss the flow generated by the boundary modular hamiltonian in the bulk. We close in section six with some discussion and open questions.

\section{Entanglement entropy, the modular hamiltonian, and relative entropy}

We consider a system that is specified by a density matrix $\rho$. This can arise in quantum 
field theory by taking a global state and reducing it to a subregion $R$. We can 
compute the von Neuman entropy $S = - Tr[ \rho \log \rho]$. Due to UV divergences this 
is infinite in quantum field theory. However, these divergences are typically independent of the 
particular state we consider,  and when they depend on the state, they do so via the expectation 
value of an operator.  See \cite{Casini:2013rba,Casini:2014yca}.

\subsection{ Modular Hamiltonian}

It is often useful to define the modular hamiltonian $K_{\rho} \equiv - \log \rho$. 
From its definition, it is not particularly clear why this is  useful -- it is in general a very non-local complicated operator. However, for certain symmetric situations it is nice and simple. 

The simplest case is a thermal state where $K = H/T$, with $H$ the Hamiltonian of the system. 
Another case is when the subregion is the Rindler wedge and the state is the vacuum of Minkowski space, 
when $K$ is the boost generator. This is a simple integral of a local operator, the stress tensor. 
For a spherical region in a conformal field theory, we have a similarly simple expression, which 
is obtained from the previous case by a conformal transformation  \cite{Casini:2011kv}. 
In  free field theory  one can also obtain a relatively 
 simple expression that is bilocal in the fields  \cite{Casini:2009sr} 
for  a  general subregion of the vacuum state. 
 
In this paper we consider another case in which simplification occurs. We consider a quantum system with a gravity dual and
a state that can be described by a  gravity solution. We will argue that  the modular Hamitonian is given
by the area of the Ryu-Takayanagi minimal surface plus the bulk modular Hamiltonian of the bulk region 
enclosed by the Ryu-Takayanagi surface.

\subsection{ Relative entropy}

Modular Hamiltonians also appear in the relative entropy
\begin{equation}
S_{rel}(\rho|\sigma)=\text{tr} \rho (\log \rho-\log \sigma)=\Delta \langle K_{\sigma} \rangle-\Delta S
\end{equation}
where $K_\sigma = - \log \sigma $ is the modular Hamiltonian associated to the state $\sigma$. 
If $\sigma$ was a thermal state, the relative entropy would be the free energy difference relative to the thermal state. 
As such it should always be positive. 

Relative entropies have a number of interesting properties such as positivity and monotonicity   \cite{Ohya.Petz}. 
Moreover, while the entanglement entropy is not well defined for QFT's, relative entropies have a precise mathematical definition  \cite{Araki:1976zv}. 

If $\rho = \sigma + \delta \rho $, then, because of positivity, the relative entropy is zero to first order in $\delta \rho$. This is called the first law of entanglement:
\begin{equation} \la{FirstLaw}
\delta S=\delta \langle K_{\sigma} \rangle
\end{equation}

When we consider a gauge theory, 
the definition of entanglement entropy is ambiguous. 
If we use the lattice definition, 
there are different operator algebras that can be naturally associated with a region $R$ 
 \cite{Casini:2013rba}.
 Different choices give different entropies. These algebras differ in the elements that are kept when splitting space into two, so that ambiguities are localized on the boundary of the region, $\partial R$.  
 One natural way of defining the entanglement entropy is by fixing a set of boundary conditions and summing over all possibilities, since there is no physical boundary. This was carried out for gauge fields in  \cite{Donnelly:2014fua, Donnelly:2015hxa, Huang:2014pfa} and gives the same result as the euclidean prescription of   \cite{Kabat:1995eq}. 
However, the details involved in the definition of the subalgebra are 
localized on the boundary.  Because of the monotonicity 
of relative entropy, these do not contribute to the relative entropy
 (see section $6$ of   \cite{Casini:2013rba} for more details).

In the case of gravitons we expect that similar results should hold. 
We expect that we similarly need to fix some boundary conditions and then sum over these choices. 
For example, we could choose to fix the metric fluctuations on the Ryu-Takayanagi surface, viewing it as a classical variable, and then 
integrate over it.
 As argued in  \cite{Casini:2013rba}, we expect that the detailed choice should not
matter when we compute the relative entropy. See appendix A for more details. 


As we mentioned above, it often occurs that two different possible definitions of the entropy give results that differ  by the expectation value of a local operator, $S(\rho)=\tr \left(\rho \mathcal{O}\right)+\tilde{S}(\rho)$.  A trivial example is the divergent area term which is just a number. 
In these cases the two possible modular Hamiltonians are related by 
\be \label{Operator}
  S(\rho)={\rm tr}\left( \rho {\cal O}\right)+\tilde{S}(\rho) 
\longrightarrow  K =   {\cal O} + \tilde K 
 \ee
 This implies that relative entropies are unambiguous,   $S(\rho|\sigma)=\tilde{S}(\rho|\sigma)$.   
For the equality of relative entropies, it is not necessary for ${\cal O}$ to be a state independent operator.
It is only necessary that ${\cal O}$ is the same operator for the states $\rho$ and $\sigma$.\footnote{
In other words, if we consider a family of states, with $\rho$ and $\sigma$ in that family, then ${\cal O}$ should be a state independent operator within
that family. }

\section{Gravity dual of the modular hamiltonian}

A leading order  holographic prescription for computing entanglement entropy was proposed in
\cite{Ryu:2006bv,Hubeny:2007xt}   and it was extended to the next order in  $G_N$  in  \cite{Faulkner:2013ana}
 (see also \cite{Barrella:2013wja}). 
The entanglement entropy of a region $R$ is the area of the extremal codimension-two surface ${\cal S}$ that asymptotes to the boundary of the region
$\partial R$, plus the bulk von Neuman entropy of the region enclosed by $\cal S $, denoted by $R_b$.  See figure \ref{RTBasic}. 
\begin{equation} \label{Entropy}
S_{\rm bdy}(R)=\frac{A_{\rm ext}({\cal S})}{4 G_N}+S_{ \rm bulk}(R_b) + S_{\rm Wald-like}
\end{equation}
$S_{\rm Wald-like}$ indicates terms which can be written as expectation values of 
local operators on $\cal S$. They arise when we compute quantum corrections  \cite{Faulkner:2013ana}, we discuss examples below.

We can extract a modular Hamiltonian from this expression. We consider states that can be described by quantum field theory in the bulk. We consider a 
reference state $\sigma$, which could be the vacuum or any other state that has a semiclassical bulk description. We consider other states $\rho$ which 
likewise can be viewed as semiclassical states built around the bulk state for $\sigma$. To be concrete we consider the situation where the classical or quantum 
fields of $\rho$ are a small perturbation on $\sigma$ so that the area is only changed by a small amount. 
Now the basic and simple observation is that both the area term and the $S_{\rm Wald-like}$ are expectation values of  operators in the bulk effective theory. 
Therefore, for states that have a bulk effective theory, we can use \nref{Operator} to conclude that 
\begin{equation}
K_{\rm bdy}=\frac{\hat{A}_{\rm ext} }{4 G_N}+\hat S_{\rm Wald-like}+K_{\rm bulk} \label{eq:modham}
\end{equation}
This includes the contribution from the gravitons, as we will explain in detail  below. The area term was first discussed in \cite{Jafferis:2014lza}. We view the area of the extremal surface as an operator in the 
bulk effective theory. This contains both the classical area as well as any changes in the area that result from the backreaction of quantum effects. 
Since we are specifying the surface using the extremality condition, this area is a gauge invariant observable in the gravity theory.\footnote{
If we merely define a surface by its coordinate location in the background solution, then a pure gauge
fluctuation of the metric can change the area. If the original surface is not extremal this already happens to 
first order. }
Note that the area changes as we change the state, but we can choose a gauge where the position of the extremal surface is fixed. Finally $\hat S_{\rm Wald-like}$ 
are the operators whose expectation values give us $S_{\rm Wald-like}$. 
 
Interestingly, all terms that can be written as local operators drop out when we consider the relative entropy. 
The relative entropy has a very simple expression
\begin{equation} \la{RelativeE}
S_{\rm bdy}(\rho | \sigma)=S_{\rm bulk}(\rho | \sigma)  
\end{equation}
Note that the term going like $1/G_N$ cancels out and we are only left with terms of order $G_N^0$. 
There could be further
corrections proportional to $G_N$ which we do not discuss in this article. It is 
tempting to speculate  that perhaps  \nref{RelativeE} might  be  true to all orders in the $G_N$ expansion (i.e. to all orders within bulk perturbation theory).

Of course, using the equation for the entropy  \nref{Entropy} and \nref{eq:modham} we can check
that the first law \nref{FirstLaw}  is obeyed. In the next section we discuss this in more detail for a spherical subregion in the vacuum.

\section{Regions with a local boundary modular Hamiltonian} 
\label{LocalK}

For thermal states,   Rindler space, or spherical regions of conformal field theories we have an explicit expression for the 
boundary modular Hamiltonian. In all these cases there is a continuation to Euclidean space with a compact euclidean time and a $U(1)$ translation symmetry along
Euclidean time. We also have a corresponding symmetry in Lorentzian signature generated by a Killing or (conformal Killing) vector $\xi$. 
 The modular Hamiltonian is then given in terms of the stress tensor as 
 $K_{\rm bdy}={\cal E}_R \equiv \int  * (\xi.T_{\rm bdy})$, where the integral is over a boundary space-like slice.  
When the theory has a gravity dual, the bulk state  $\sigma$ is also invariant under a bulk Killing vector $\xi$. 
In this subsection we will discuss  \nref{eq:modham} for  states constructed around $\sigma$.

For this discussion it is useful to recall Wald's treatment of the first law \cite{Bardeen:1973gs,Wald:1993nt,Iyer:1994ys} 
\begin{equation} \la{WaldFirst}
\delta {\cal E}_R= { A_{\rm lin}(\delta g) \over 4 G_N}  + \int_{\Sigma} * ( \xi.E_g(\delta g))
\end{equation} 
 where $E_g(\delta g)$ is simply the linearized  Einstein tensor   with the proper cosmological constant. 
  It is just the variation of the gravitational part of the action and does not include the matter contribution. 
  Here $A_{\rm lin}$ is the first order variation in the area due to a metric fluctuation $\delta g$.
   And $\Sigma$ is any  Cauchy slice in the entanglement wedge $R_b$.
 Equation \nref{WaldFirst} is a tautology, it arises by integrating by parts the linearized Einstein tensor. 
 It is linear in $\delta g$ and we can write it as an operator equation by sending $\delta g \rightarrow \delta \hat g $,
  where $\delta \hat g$ is the operator describing
 small fluctuations in the metric in the semiclassically quantized theory.

%
 
\subsection{ Linear order in the metric}

For clarity we will first ignore dynamical gravitons, and include them later (we would have nothing extra to include if we were in three bulk dimensions). 
We  consider   matter fields with an $o(G_N^0)$ stress tensor in the bulk, assuming the matter stress tensor was zero on the $\sigma$ background.\footnote{This discussion can be simply extended when there is a non-zero but $U(1)$-symmetric background matter stress tensor, such as in a charged black hole. In that case we need to subtract the background stress tensor to obtain the bulk modular Hamiltonian.} 
Such matter fields produce a small change in the metric that can be obtained
by linearizing the Einstein equations around the vacuum. These equations say 
$E_g(\delta g)_{\mu \nu}  = T_{\mu \nu}^{\rm mat} $, where $T_{\mu \nu}^{\rm mat}$ is the stress tensor of matter. Inserting this in \nref{WaldFirst} we find that 
\cite{Bardeen:1973gs,Wald:1993nt,Iyer:1994ys} 
\be
\delta {\cal E}_R =   { A_{\rm lin} (\delta g) \over 4 G_N}+ \int_{\Sigma} * ( \xi . T)  =  {  A_{\rm lin}(\delta g ) \over 4 G_N}+ K_{\rm bulk } \la{AreaLin}
\ee
where we used that the bulk modular Hamiltonian also has a simple local expression 
in terms of the stress tensor  due 
to the presence of a Killing vector with the right properties at the entangling surface $\cal S $. 
  Notice that we can disregard additive constants in both the area and ${\cal E}$, which are the values for the
state $\sigma$. We only care about deviations from these values.  
This is basically the inverse of the argument in \cite{Swingle:2014uza}.  This shows how \nref{eq:modham} works in this symmetric case.
The term $\hat S_{\rm Wald-like}$ in \nref{eq:modham} 
 arises in some cases as we discuss below. 

Let us now discuss the $ \hat S_{\rm Wald-like}$ term. There can be different sources for this term. 
A simple source  is the following.
The bulk entanglement entropy has a series of divergences which include an area term, but also terms with higher powers of the curvature. Depending on how
we extract the divergences we can get certain terms with finite coefficients.  Such terms 
are included in $S_{\rm Wald-like}$. A different case is that of a scalar field with a coupling 
$\alpha \phi^2(R -R_0)$ where $R$ is the Ricci scalar in the 
bulk, and $R_0$ the Ricci scalar on the unpertubed background, the one associated to the state $\sigma$. Then there exists an additional term in the entropy of the form $\hat S_{\rm Wald-like} = 2 \pi \alpha  \int_{\cal S} \phi^2 $. 
   If we compute the entropy  as the continuum limit of the one  on the lattice, then it will be independent of $\alpha$.  
Under these conditions the bulk modular Hamiltonian is also
 independent of 
$\alpha$ and  is given by the canonical stress tensor, involving only first derivatives of the field. However, the combination of 
$K_{\rm bulk} + \hat S_{\rm Wald-like}  = \int_{\Sigma} * ( \xi . T^{\rm grav}(\phi))$, where $T^{\rm grav}_{\mu \nu}(\phi)$
 is the standard stress tensor that would appear in the right hand
side of Einstein's equations. $T^{\rm grav}_{\mu \nu}(\phi)$  does depend on $\alpha$. The $\alpha $ dependent contribution is a total derivative which evaluates to $ 2 \pi \alpha \phi^2$ at
the extremal surface. A related discussion in the field theory context appeared in   \cite{Casini:2014yca,Lee:2014zaa}.  



\subsection{ The  graviton contribution} 

We expect that we can view the propagating gravitons as one more field that lives on the original background, 
given by the metric $g_\sigma$. 
In fact,  we can  expand Einstein's equations in terms of $g = g_\sigma + \delta g_2  + h$. Here $h$, which is of order $\sqrt{G_N}$, represents the 
dynamical graviton field and obeys linearized field equations. $\delta g_2$ takes into account the effects of back-reaction and obeys
the equation 
\be
E(\delta g_2)_{\mu \nu} = T^{\rm grav}_{\mu \nu}(h) + T^{\rm matter}_{\mu \nu } 
\ee
where $T^{\rm grav}_{\mu\nu}(h) $ comes simply from expanding the Einstein tensor (plus 
the cosmological constant) to second order and moving the quadratic term in $h$ to the right hand side. 
 $h$ obeys the homogeneous linearized equation of motion, so the term linear in $h$ in the equation above vanishes. 
We can now use equations (44-46) in \cite{Hollands:2012sf}, which imply that 
\be \la{Express}
K_{\rm bdy, 1 + 2} = {\cal E}_{1+2}=  { \hat A_{\rm lin}(h + \delta g_2) + \hat A_{\rm quad}(h) \over 4 G_N} +  { E}_{\rm can} 
\ee
where $K_{\rm bdy, 1 + 2}$ is the boundary  modular Hamitonian (or energy conjugate to $\tau$ translations) expanded
to quadratic order in fluctuations. Similarly, the area is expanded to linear and quadratic order. Finally, ${ E}_{\rm can}$ is the bulk canonical energy\footnote{This differs from the integral of the gravitational stress tensor by boundary terms.} defined by 
${ E}_{\rm can } = \int \omega( h , {\cal L}_\xi h ) + $matter contribution, where $\omega$ is the symplectic form defined in 
\cite{Hollands:2012sf}. From this expression we 
conclude that the modular Hamiltonian is the canonical energy 
 \be \la{KeqE}
 K_{\rm bulk} = { E}_{\rm can} 
 \ee
 
We can make contact with the previous expression \nref{AreaLin} as follows. If we include the gravitons 
by replacing  $T^{\rm mat}_{\mu \nu} \to T_{\mu\nu}^{\rm mat} + T^{\rm grav}_{\mu \nu}(h)$  in 
\nref{AreaLin}, then we notice that we  get $A_{\rm lin}(\delta g_2)$, without the
term $A_{\rm quad} (h)$. However, one can argue that (see eqn. (84) of \cite{Hollands:2012sf})
\be
 \int_\Sigma * ( \xi . T^{\rm grav} (h) ) = { E}_{\rm can} (h) + { A_{\rm quad}(h) \over 4 G_N} 
\ee
thus recovering \nref{Express}. 

In appendix A we discuss in more detail the  boundary conditions that are necessary for quantizing the
graviton field. 
 
\subsection{ Quadratic order for coherent states}

The problem of the gravity dual of relative entropy was considered in 
 \cite{Lashkari:2015hha} in the classical regime for quadratic fluctuations around a background with a local 
modular Hamiltonian. They argued that the gravity dual is equal to the canonical energy. 
Here we rederive their result from \nref{RelativeE}. 

We simply view a classical background as a coherent state in the quantum theory. 
  $e^{i \lambda \int \Pi \hat{\phi}+\phi \hat{\Pi}}|\psi_\sigma \rangle$, where $|\psi_\sigma\rangle$ is the state associated to $\sigma$ \footnote{ 
Here  $\lambda$ could be $O(1/\sqrt{G_N})$  as long as the backreaction is small.  }. 
   We see that in free field theory we can view coherent states as arising from the action of a product of unitary operators, one acting inside the region and one ouside.   
 For this reason finite coherent excitations do not change the 
bulk   von Neuman entropy of subregions, or  $\Delta S_{\rm bulk }=0$.  Thus, the contribution to the bulk relative entropy comes purely from the bulk Hamiltonian, which we have argued is equal
  to canonical energy \nref{KeqE} . Therefore, in this situation we recover  the result in \cite{Lashkari:2015hha}
  \be 
  S_{\rm bdy}(\rho |\sigma) = S_{\rm bulk}(\rho|\sigma) = \Delta K_{\rm bulk} - \Delta S_{\rm bulk } = \Delta K_{\rm bulk} = E_{\rm canonical} 
  \ee

\section{Modular flow} 

The modular hamiltonian generates an automorphism on the operator algebra, the modular flow. Consider the unitary transformation $U(s)=e^{i K s}$. Even if the modular hamiltonian is not technically an operator in the algebra, the modular flow of an operator, $O(s) \equiv U(s) O U(-s)$, stays within the algebra. 
For a generic region,  the modular flow might be complicated, see \cite{Casini:2009vk} for some discussion about modular flows for fermions in  $1+1$ dimensions. 
However, in our holographic context it can help us understand subregion-subregion duality. In particular, it can help answer 
 the question of whether the boundary region $R$ 
 describes the entanglement wedge or only the causal wedge  \cite{Almheiri:2014lwa,Czech:2012bh,Bousso:2012sj,Headrick:2014cta}. The entanglement wedge is the causal domain of the spatial region bounded by the interior of ${\cal S}$.
 
From \nref{ModHam}, we have that
\be \label{modflow}
[ K_{\text{bdy}},\phi]=[K_{\text{bulk}},\phi] 
\ee
where $\phi$ is any operator with support only in the interior of the entanglement wedge, and where on the right-hand side we have suppressed terms subleading in $G_N$. On the left-hand side terms in $K_{\rm bdy}$ localized on ${\cal S}$ have dropped out, similarly as in \nref{RelativeE}. Thus the boundary modular flow is equal to the bulk modular flow of the entanglement wedge,
the causal wedge does not  play any role.

One may also consider the flow generated by the total modular operator, 
$K_{\rm bdy, Total}=K_{{\rm bdy}, R}-K_{{\rm bdy}\bar R}$, which should be a smooth operator without any ambiguities. From our full formula for the bulk dual of the modular Hamiltonian we see that 
 $K_{\rm bdy, \rm Total}=K_{\rm bulk, Total} + o(G_N)$.  If the global state is pure, then $K_{\rm Total}$ annihillates it. 



\subsection{Smoothness of the full modular Hamiltonian in the bulk}

For problems that have a $U(1)$ symmetry, such as thermal states and Rindler or spherical subregions of CFTs, we know the full boundary  modular Hamiltonian ${\cal E}$. We can define a time coordinate $\tau$ which is translated by the action of ${\cal E}$ in the boundary theory.   
In these situations the bulk state also has an associated  symmetry generated by the Killing vector $\xi$. 
We can choose coordinates so that we extend $\tau$ in the bulk and $\xi$ simply translates
$\tau$ in the bulk. Then the bulk modular Hamiltonian is the bulk operator that performs a translation of the bulk fields along the bulk $\tau $ direction. 

\begin{figure}[h]
\begin{center}
\vspace{5mm}
\includegraphics[scale=1]{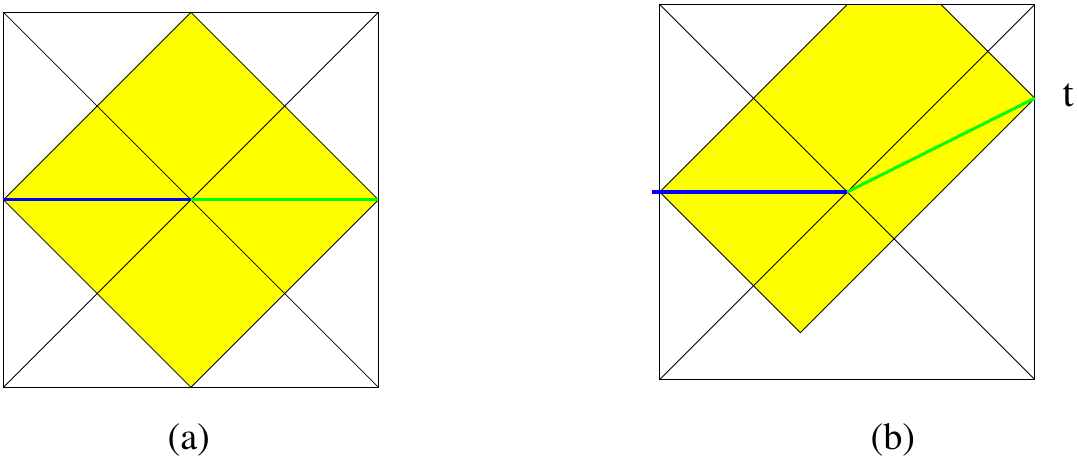}
\vspace{5mm}
\caption{ In this figure we are considering the thermofield double state. (a) Acting with the bulk modular Hamiltonian 
$e^{- i  t K_{\rm  bulk,R }}$ we get a new state on the horizontal line that has a singularity at the horizon. (b) The area term introduces a kink, or a relative boost between the left and right sides.  Then  the state produced by the full right side Hamiltonian is non-singular, and locally equal to the vacuum state. 
  }
\label{modhamaction}
\end{center}
\end{figure}

Let us now consider an  eternal black hole and the thermofield double state \cite{Maldacena:2001kr}. 
This state is invariant under the action of $H_R-H_L$. Let us now consider the action of only the
right side boundary Hamiltonian $H_R$\footnote{Here left and right denote the two copies in the thermofield double state.}.
It was argued in  \cite{Maldacena:2013xja} that  this corresponds to the same gravity solution but where the origin of the time direction on the right side is changed. This implies that the Wheeler de Witt patch associated to $t_L = t_R=0$ looks as in figure \ref{modhamaction}(b), after the action of $e^{- i t H_R}$
On the other hand, if we consider the bulk quantum field theory and we act with only the 
right side bulk modular Hamiltonian $K_{\rm bulk , R}$ we would produce a state that is singular at 
the horizon. By the way, it is precisely for this reason that algebraic quantum field theorists like to consider the total modular Hamiltonian instead. 
It turns out that  the change in the bulk state is the same as the one would obtain if we were quantizing
the bulk field theory along a slice which had a kink as shown in figure \ref{modhamaction}(b). 
Interestingly the area term in the full modular Hamiltonian \nref{eq:modham} has the effect of producing
such a kink. In other words,  the area term produces a shift in the $\tau$ coordinate, or a relative boost
between the left and right sides \cite{Carlip:1993sa}. The action of only the area term or only 
$K_{\rm Bulk, R}$ would lead to a state that is singular at the horizon, but the combined action of the
two produces a smooth state, which is simply the same bulk geometry but with a relative shift in the 
identification of the boundary time coordinates\footnote{We thank D. Marolf for discussions about this point.}. 
 
Let us go  back to a  general non-U(1) invariant case.  
Since the bulk modular Hamiltonian reduces to the one in the $U(1) $-symmetric case very near 
 the bulk entangling surface ${\cal S}$, we expect that the action of the full boundary modular 
Hamiltonian, including the area term, will not be locally singular in the bulk -- though it can be singular from the boundary point of view due to boundary UV divergences.

\subsection{Implications for entanglement wedge reconstruction} \label{sec: EW}

One is often interested in defining local bulk operators as smeared operators in the boundary. This operator should be defined order by order in $G_N$ over a fixed background and should be local to the extent allowed by gauge constraints. If we consider a $t=0$ slice in the vacuum state, then we can think of a local bulk operator $\phi(X)$ as a smeared integral of boundary operators  \cite{Hamilton:2006az}
\begin{equation}
\phi(X)=\int_{bdy} dx^{d-1} dt\, G(X|x,t) O(x,t)+o(G_N)
\end{equation} One would like to understand to what extent this $\phi$ operator can be localized to a subregion in the boundary.

Given a region in the boundary $R$, we have been associating a corresponding region in the bulk, the so-called entanglement wedge which is the domain of dependence of $R_b$, $D[R_b]$. There is another bulk region one can associate to $R$, the causal wedge (with space-like slice $R_C$) which is the set of all bulk points in causal contact with $D[R]$, \cite{Hubeny:2012wa}. $R_C$ is generically 
smaller than $R_b$  \cite{Wall:2012uf, Headrick:2014cta}.  
 
In situations with a $U(1)$ symmetry, such as a thermal state or a Rindler or spherical subregion of a CFT, we have time-translation symmetry and a local modular Hamiltonian that generates translations in the time $\tau$.  We 
can express bulk local operators in the entanglement wedge (which coincides with the causal wedge) in terms of boundary operators localized in $D[R]$ \cite{Hamilton:2006az,Morrison:2014jha}\footnote{It is sometimes necessary to go to Fourier space to make this formula precise \cite{Morrison:2014jha,Papadodimas:2012aq}.} 
 %
%
\begin{equation} \la{Usual}
\phi(X)=\int_{R} d y^{d-1} \int  d\tau\,  G'(X|y,\tau) O(y,\tau )+o(G_N)\ , \qquad X \in R_b
\end{equation} 
 
A natural proposal for describing operators in that case is that we can replace $\tau$ in \nref{Usual} by the modular parameter $s$. In other words, we consider
modular flows of local operators on the boundary, defined as $O_R(x,s) \equiv U(s) O_R(x, 0) U^{-1}(s)$

\begin{figure}[h]
\begin{center}
\vspace{5mm}
\includegraphics[scale=2]{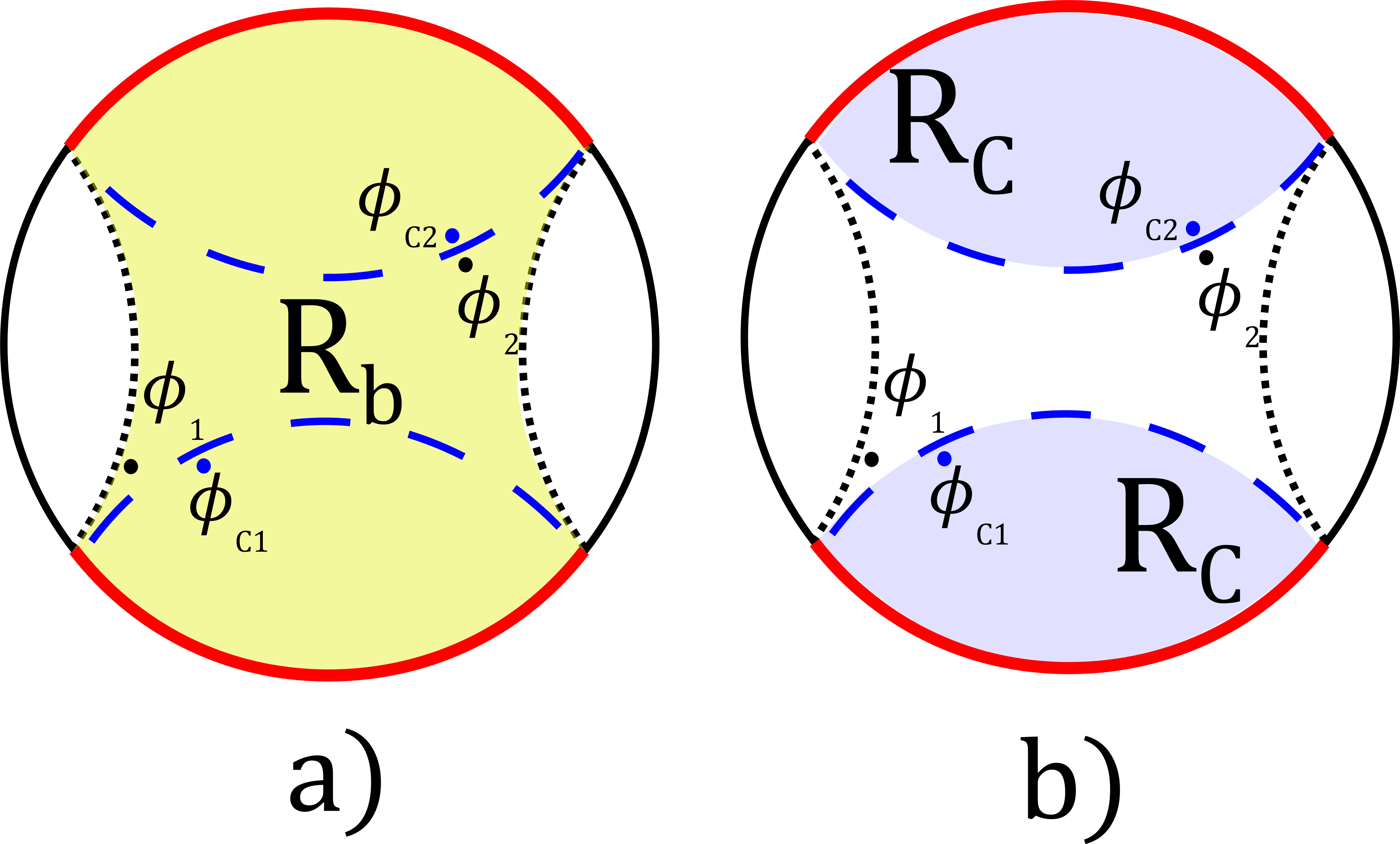}
\vspace{5mm}
\caption{  In both figures the region $R$ is the union of the two red intervals and the Ryu-Takayanagi
 surface is the dotted black line, while the boundary of $R_C$ is the blue dashed line (color online). 
In $a)$, the  shaded region denotes the defining spatial slice $R_b$ of the entanglement wedge. 
 In $b)$, the shaded region is the defining spatial slice $R_C$ of the causal wedge.
The modular flow of an operator close to the Ryu-Takayanagi
  surface will be approximately local, so that $\phi_1(s)$ will be almost local and, after some $s$,
 it will be in causal contact with $\phi_{C1}$. 
This flow takes the operator out of this slice to its past or to its
 future.  Alternatively, if we consider an operator near the boundary of the causal wedge $\phi_{C2}$, it is clear that, under modular flow, $[\phi_{C2}(s),\phi_2] \not=0$.    }
\label{ewedge}
\end{center}
\end{figure}

 A simple case in which $R_b$ is larger than $R_C$ is the case of two intervals in a 1+1 CFT such that their total size is larger than half the size of the whole system, see figure \ref{ewedge}. Here, it is less clear how to think about the operators in the entanglement wedge.  We would like to use the previous fact that the modular flow is bulk modular flow to try to get some insight into this issue.

The modular flow in the entanglement wedge will be non-local, but highly constrained: the bulk modular hamiltonian is bilocal
 in the fields  \cite{Casini:2009sr}.
 If we have an operator near the boundary of the causal wedge and modular evolve it, it will quickly develop a non zero commutator with a nearby operator which 
 does not lie in the causal wedge. Alternatively, an operator close to the boundary of the entanglement wedge will have an approximately local modular flow. It will follow the light rays emanating from the extremal surface and it can be on causal contact with the operators in the causal wedge.  See figure \ref{ewedge}.

So we see that to reconstruct the operator in the interior of the entanglement wedge, one necessarily needs to understand better  the modular flow. It seems natural to conjecture that one can generalize \nref{Usual}
 to two intervals (or general regions) 
 by considering the modular parameter instead of Rindler time, ie the simplest generalization of the AdS/Rindler formula which accounts for the non-locality of the modular hamiltonian would be  \begin{equation}
\phi(X)=\int_R dx \int ds G''(X| x,s) O(x,s)\ , \qquad X \in R_b
\end{equation} 
Here $G''$ is a function that should be worked out. It will depend on the bilocal kernel that describes the modular Hamiltonian for free fields  \cite{Casini:2009sr}.

So we see that to reconstruct the operator in the interior of the entanglement wedge, it is necessary to understand better modular flows in the  quantum field theory of the bulk. 
 To make these comments more precise, a more detailed analysis would be required, which should include a discussion about gravitational dressing and the constraints. We leave this to future work.

 Here we have discussed how the operators in the entanglement wedge can be though of from the boundary perspective. However, note that from \nref{eq:modham} (and consequently the formula for the relative entropy), it is clear that one should think of the entanglement wedge as the only meaningful candidate for the ``dual of $R$", see also \cite{Almheiri:2014lwa}. If we add some particles to the vacuum in the entanglement wedge $R_b$ (which do not need to be entangled with $\bar{R}_b$), the bulk  relative entropy will change. According to \nref{RelRel}, the boundary relative entropy also changes and, therefore,   state is distinguishable from the vacuum, even
 if we have only access to $R$.

\section{Comments and discussion }

\subsection{The relative entropy for coherent states}
 
If we consider coherent states, since their bulk  entanglement entropy is not changed, the relative entropy will just come from the difference in the bulk modular hamiltonian. Since our formulation is completely general, one could in principle compute it for any reference region or state and small perturbations over it.

A particularly simple case would be the relative entropy for an arbitrary subregion between the vacuum and a coherent state of matter. To second order in the perturbation, one only needs to work out how the modular hamiltonian for the free fields \cite{Casini:2009sr} looks like for that subregion of AdS, and then evaluate it in the coherent state background.

\subsection{Positivity of relative entropy and energy constraints}

Our formula \nref{RelativeE}  implies  that the energy constraints obtained from the positivity of the relative entropy 
can be understood as arising from the fact that the relative entropy has to be positive in the bulk.

\subsection{Higher derivative gravity}

Even though we focused on Einstein gravity, our discussion is likely to apply to other theories of gravity. The modular hamiltonian will likely be that of  an operator localized on the entangling surface plus the bulk modular Hamiltonian in the corresponding entanglement wedge. Thus the relative entropy will be that of the bulk. 
There could be subtleties that we have not thought about.


\subsection{Beyond extremal surfaces }

A. Wall  proved the second law by using the monotonicity of 
relative entropy \cite{Wall:2010cj,Wall:2011hj}. If we consider 
 two Cauchy slices $\Sigma_0,\Sigma_{t>0}$ outside a black hole, then $S_{rel,t}<S_{rel,0}$ is enough to prove the generalized
second law.  Interestingly,  section 3 of \cite{Hollands:2012sf} shows the ``decrease of canonical energy": ${ E_{\rm can}}(t)<{ E_{\rm can}}(0)$. The setup (Cauchy slices) that they both consider is the same. 
Due to the connection between relative entropy and canonical energy,  \cite{Lashkari:2015hha}, we
expect a relation between these two statements. This does not obviously follow from what we said due to the following reason. 

Here we   
 limited  our discussion to the entanglement wedge. 
In other words, we are always considering the surface $\cal S $ to be extremal.   We expect that the discussion should generalize to situations where the surface $\cal S$ is along a causal horizon. The question is: what 
is the precise boundary dual of the region exterior to such a horizon? Even though we can think about the bulk computation, we are not sure what
boundary computation it corresponds to. A proposal was made in \cite{Kelly:2013aja},
 and perhaps one can understand it in that context. 


Being able to define relative entropies for regions which are not bounded by minimal surfaces is also 
crucial to the interesting proposal  in \cite{Jacobson:2015hqa} to derive Einstein's equations from (a suitable extension to non-extremal surfaces of) the Ryu-Takayanagi formula for entanglement.

\subsection{Distillable entanglement}

In the recent papers \cite{Soni:2015yga, VanAcoleyen:2015ccp} it was argued that for gauge fields, only the purely quantum part of the entanglement entropy
corresponds to distillable entanglement. The ``classical" piece that cannot be used as a resource corresponds to the shannon entropy of the center variables of \cite{Casini:2013rba}. Our terms local in $\cal S$ are the gravitational analog of this classical piece and one might expect that a bulk  observer with access only to the low-energy effective field theory can only extract bell pairs from the bulk entanglement. This seems relevant for the AMPS paradox \cite{Almheiri:2012rt,Almheiri:2013hfa,Marolf:2013dba}. 


\section*{Acknowledgments}

We thank H. Casini, M. Guica, D. Harlow, T. Jacobson, N. Lashkari, J. Lin, 
D. Marolf,  H. Ooguri, M. Rangamani, and A. Wall for discussions. The work of D.L.J. is supported in part by NSFCAREER grant PHY-1352084. A.L. was supported in part by the US NSF under Grant No.~PHY-1314198. 
J.M. was supported in part   by U.S. Department of Energy grant
de-sc0009988. J.S. was supported at KITP in part by the National Science Foundation under Grant No. NSF PHY11-25915, and is now supported in part by the Natural Sciences and Engineering Research Council of Canada and by grant 376206 from the Simons Foundation.

\appendix


 

\section{Subregions of gauge theories} 

\subsection{U(1) gauge theory}

The problem of defining the operator algebra of a subregion of a gauge theory was considered in \cite{Casini:2013rba}. 
It was shown that for a lattice gauge theory there are several possible definitions of the subalgebra. It was further found that
the subalgebra can have  a center, namely some operators that commute with all the other elements of the subalgebra. In this case 
we can view the center as classical variables. Calling the classical variables $x_i$, then for each value of $x_i$ we have a classical probability $p_i$ and a 
density matrix $\rho_i$ for  each irreducible block.  The relative entropy between two states is then 
\be
 S(\rho |\sigma ) = H(p |q)  + \sum_i p_i S(\rho_i |\sigma_i ) 
 \ee
 where $p_i, ~q_i$ are the probabilities of variables $x_i$ in the state $\rho$ and $\sigma$ respectively. 
 $H$ is the classical (Shanon) relative entropies of two probability distributions, $H = \sum_i p_i \log( p_i/q_i) $. 

In the continuum we expect that the relative entropy is finite and independent of the microscopic details regarding the precise definition of the
algebra \cite{Araki:1976zv}. 

These microscopic details have a continum counterpart. When we consider a region $R$ we would like to be able to define a consistent quantum theory 
within the subregion. In particular, imagine that we consider all classical solutions restricted to the subregion. Then we define a presymplectic product between 
two such solutions, which we will use to quantize the gauge orbits. 
This presymplectic product should be gauge invariant so that it does not depend on the particular representative. 
Let us consider a free Maxwell field. The presymplectic product is given by integrating  
\be 
 \Omega(A^1,A^2)=\int_{\Sigma} \omega(A^1,A^2) = \int_{\Sigma}  (A^1 \wedge *F^2 - A^2 \wedge *F^1) 
 \ee
 where $A^1 = A^1_\mu dx^\mu$ is a gauge field configuration. Here we imagine that both $A^1$ and $A^2$ are solutions to the equations of motion. 
 $\Sigma$ is any spacelike surface. 
   
 Demanding gauge invariance amounts to the statement 
 \be
 0=\Omega(A , d\epsilon) = \int_{\partial \Sigma   } \epsilon \wedge F
\ee
where $ \partial \Sigma$ is the boundary of the spacelike surface.
 We have used the equations of motion for $F$ and integrated by parts. 
In order to make this vanish we need some boundary conditions. In particular, let us concentrate on the boundary conditions  required at the 
boundary of $\Sigma $ corresponding to the boundary of a region  ${\cal S} = \partial \Sigma $.
One possible boundary condition is to set $A_i=A^{\rm cl}_i$ for components along the surface, where $A^{\rm cl}_i$ is a classical gauge field on the surface. 
In this case, it is natural to set $\epsilon =0$ on the surface. We can quantize the problem for each fixed $A^{\rm cl}_i$ and then integrate over all $A^{\rm cl}_i$. 
These values of $A^{\rm cl}_i$ are the ``center'' variables $x_i$ in the above discussion. This is called the ``magnetic'' center, since the gauge field $A^{\rm cl}_i$ defines 
a magnetic field $F = dA^{\rm cl} $ on the surface. 

There are other possibilities, such as fixing the electric field, or ``electric center'', where the perpendicular electric field is fixed. 

These would correspond to specific choices on the lattice. Since we expect that relative entropy is a finite and smooth function of the shape of the region, \cite{Casini:2013rba}  has shown that the detailed boundary condition does not matter, as long as we choose something that makes physical sense.  Recently, \cite{Donnelly:2014fua,Donnelly:2015hxa} carried out explicitly the field theory calculation, being careful with the center variables.
 
  \subsection{Gravity } 
  
   Here we consider the problem of defining a subregion in a theory of Einstein gravity. We consider only the problem at the quadratic level where we need
   to consider free gravitons moving around a fixed background (which obeys Einstein's equations). 
   These gravitons can be viewed as a particular example of a gauge theory. We can also compute the symplectic form, as given in \cite{Iyer:1994ys}, and 
   then impose that the symplectic inner product between a pure gauge mode and another solution to the linearized equations vanishes. 
   Here the gauge tranformations are reparametrizations, generated by a vector field $\zeta$.
Note that $\zeta$ is {\it not} a killing vector, it is a general vector field and
 it should not be confused with $\xi$ discussed in section \ref{LocalK}.  Writing the metric as $ g+ \delta g$, where $g$ is the background metric and
   $\delta g $ is a small fluctuation. Then the gauge tranformation acts as  $\delta g \to \delta g + { \cal L}_{\zeta} g $, where ${\cal L}_\zeta $ is the Lie derivative. 
   Then, as shown in \cite{Hollands:2012sf}, there is a simple expression for the sympectic product with a such a pure gauge mode 
    \be \la{Symp}
  \int_{\Sigma} \omega( \delta g , {\cal L}_\zeta g ) = \int_{\partial \Sigma} \delta  Q_\zeta - \zeta . \Theta(g,\delta g) 
   \ee
   with $Q_\zeta$ and $\Theta(g,\delta g)$ given in  eqns (32) and (17) of \cite{Hollands:2012sf}.

   We would like to choose boundary conditions on the surface which make the right hand side zero. 
   We choose boundary conditions similar to the ``magnetic'' ones above. Namely, we
 fix the metric along the entangling surface $\cal S $ to $\delta g_{ij}= \gamma_{ij}$. 
   We treat $\gamma_{ij}$ as classical and then integrate over it. This is enough to make all terms in \nref{Symp} vanish. Let us be more explicit. 
   By a change of coordinates we can always set the metric to have the following form near the entangling surface. For simplicity we write it in Euclidean space, but
   the same is true in Lorentzian signature 
   \be \la{GaugeFixed}
   ds^2 =  d\rho^2 + [ \rho^2 + o(\rho^4) ]( d\tau + a_i dy^i)^2 + h_{ij} dy^i dy^j 
   \ee
   here $a_i$ and $h_{ij}$ can be functions of $\tau$ and $\rho$, with a regular expansion around $\rho=0$.
In these coordinates the extremal surface ${\cal S}$ is always at $\rho=0$, both for the original metric and
the perturbed metric.  
  Extremality implies  that the trace of the extrinsic curvature is zero,  or 
   $K^A = h^{ij} \partial_{X^A} h_{ij} =0$, where $X^A = (X^1,X^2) = (\rho \cos \tau, \rho \sin \tau ) $.
This is true for the background and the fluctuations 
   \be \la{Extr}
   K^A =0 , ~~~~~~ \delta K^A=0
   \ee
    which ensures that even on the perturbed solution we are considering the minimal surface.
These conditions ensure that the splitting between the two regions is defined in a gauge invariant way. 
   
   We demand that all fluctuations are given in the gauge \nref{GaugeFixed}. Thus, near $\rho=0$, 
$\delta g$ leads to $\delta a_i$ and $\delta h_{ij}$. 
 We now further set a boundary condition that 
   $\delta h_{ij} = \gamma_{ij}$ where $\gamma_{ij}$ is a classical function which we will later integrate
 over. For defining the quantum problem we will view it as being classical. We will quantize the fields in the subregion for fixed values of $\gamma_{ij}$ and then integrate over the classical values of $\gamma_{ij}$. 
   
    With these boundary conditions we see that all terms in \nref{Symp} vanish. In fact, \nref{Symp}, has three terms\footnote{We did not keep track of the numerical coeficients in front of each of the three terms}
    \bea \la{ThreeT}
\int_{\Sigma}\omega( \delta h , {\cal L}_\xi g ) &= &  \int_{\partial \Sigma}\delta_{\delta h}Q(\zeta) - i_\zeta \Theta(g,\delta h) \nonumber
\\
&= &\int_{\partial \Sigma} \left[  \delta a_i  \zeta^i  +  \zeta^\tau \delta h^i_i  + ( - h^{ij} \partial_A \delta h_{ij} + {1\over 2 } \delta h^{ij} \partial_A h_{ij} ) \zeta_B  \epsilon^{AB} \right]   
\eea
   
    Since the fluctuation of the metric is zero    at the entangling surface, $\delta h_{ij} =0$, we see that many terms vanish. 
    In addition, since we are setting $\delta h_{ij} =0$, it is also natural to restrict the vector fields so that $\zeta^i=0$ on the surface. 
    This ensures that the first term in \nref{ThreeT} vanishes. Note that the middle term is related to the 
fact that the area generates a shift in the coordinate $\tau$. After all the area is the Noether charge
associated to such shifts \cite{Wald:1993nt,Iyer:1994ys}.

    
    The extremality condition makes sure that we are choosing a (generically) unique surface for each geometry. We then treat the 
    induced geometry on the surface as a classical variable, quantize the metric in the subregion, and then sum over this classical variable. In this region, we
    seem to have a gauge invariant symplectic product.

    We have not explicitly computed the entanglement entropy for gravitons with these choices, but we expect that it should lead to a well defined problem and
    that relative entropies will be finite.


\end{document}